\documentclass[sn-mathphys-num]{sn-jnl}

\usepackage{graphicx}%
\usepackage{multirow}%
\usepackage{amsmath,amssymb,amsfonts}%
\usepackage{amsthm}%
\usepackage{mathrsfs}%
\usepackage[title]{appendix}%
\usepackage{xcolor}%
\usepackage{textcomp}%
\usepackage{manyfoot}%
\usepackage{booktabs}%
\usepackage{algorithm}%
\usepackage{algorithmicx}%
\usepackage{algpseudocode}%
\usepackage{listings}%

\theoremstyle{thmstyleone}%
\newtheorem{theorem}{Theorem}
\newtheorem{proposition}[theorem]{Proposition}%

\theoremstyle{thmstyletwo}%
\newtheorem{remark}{Remark}%

\theoremstyle{thmstylethree}%
\newtheorem{definition}{Definition}%

\begin{document}

\title[Article Title]{Distributed Convex Optimization with State-Dependent (Social) Interactions over Random Networks}


\author*[1]{\fnm{Seyyed Shaho} \sur{Alaviani}}\email{salavian@umn.edu}

\author[2]{\fnm{Atul} \sur{Kelkar}}\email{akelkar1@binghamton.edu}


\affil*[1]{\orgdiv{Department of Mechanical Engineering}, \orgname{University of Minnesota}, \orgaddress{\city{Minneapolis}, \state{Minnesota}, \country{USA}}}

\affil[2]{\orgdiv{Dean, Thomas J. Watson College of Engineering and Applied Science}, \orgname{Binghamton University}, \orgaddress{\city{Binghamton}, \state{New York}, \country{USA}}}



\abstract{This paper aims at distributed multi-agent convex optimization where the
	communications network among the agents are presented by a \textit{random}
	sequence of possibly \textit{state-dependent} weighted graphs. This is the \textit{first} work to consider both random arbitrary communication networks and state-dependent interactions among agents. The state-dependent weighted random operator of the graph is shown to be \textit{quasi-nonexpansive};
	this property neglects a priori distribution assumption
	of random communication topologies to be imposed on the operator. Therefore, it contains more general class of
	random networks with or without asynchronous protocols. A more general mathematical optimization
	problem than that addressed in the literature is presented, namely minimization of a convex function over the fixed-value point set of a quasi-nonexpansive random operator.
	A discrete-time algorithm is provided that is able to converge \textit{both} almost surely and in mean square to the global solution of the optimization problem. Hence, as a special case,
	it reduces to a \textit{totally asynchronous} algorithm for the
	distributed optimization problem. The algorithm is able to converge even if the weighted matrix of the graph is periodic and irreducible under synchronous protocol. Finally, a case study on a network of robots in an automated warehouse is
	given where there is \textit{distribution dependency} among random communication graphs.     }




\maketitle

\section{Introduction}\label{sec1}

Distributed multi-agent optimization has been an attractive topic due to its
applications in several areas such as power systems, smart buildings, and machine learning to name a few; therefore, several investigators have paid much attention to distributed optimization problems (see Surveys \cite{survey1}-\cite{survey4}). \textit{Switched} dynamical systems are divided into two categories: \textit{arbitrary} (or state-independent) and \textit{state-dependent} (see \cite{11222} and references therein for details and several examples). Many references, cited in Surveys \cite{survey1}-\cite{survey4}, have investigated distributed optimization over \textit{arbitrary} networks.

On the other hand, state-dependent networks have been shown in practical systems
such as flocking of birds \cite{38}, opinion dynamics \cite{36}-\cite{ozdaglar5}, mobile robotic networks \cite{33}, wireless networks \cite{34}, and predator-prey interaction
\cite{35}. For example, an agent in social networks weighs the opinions of others based on how much its opinion is close to theirs (see Section I in the preliminary version, i.e., \cite{alavianiCDC2021}, for more details).	

In state-dependent networks, coupling between algorithm analysis and information exchange among agents impose significant challenge because states of agents at each time determine the weights in the communication networks. Hence,  distributed algorithms' design for consensus and
optimization over state-dependent networks is still a challenge.  

Consensus problem for opinion dynamics has been investigated in \cite{36}-\cite{ozdaglar5}. Existence of consensus in a multi-robot network has been shown in \cite{42}. Distributed
consensus \cite{statedependentbothways}-\cite{45} and distributed optimization \cite{8}, \cite{11}-\cite{alavianiSignalprocessing} over
state-dependent networks with time-invariant or time-varying\footnote{The underlying communication graph is a priori known in a time-varying arbitrary network at each time $t$, whereas it is a priori \textit{unknown} in a random arbitrary network.} arbitrary graphs have been considered. Hence, the \textit{gap} in the literature is to consider distributed multi-agent optimization problems with both \textit{state-dependent} interactions and \textit{random} arbitrary (see footnote 1) networks.



This paper aims at distributed multi-agent convex optimization over \textit{both} state-dependent and random arbitrary
networks, that has not been addressed in the literature. Assuming doubly stochasticity of weighted matrix of the graph with respect to
state variables for each communication network and strong connectivity of the
union of the communication networks allows this result to be applicable to \textit{periodic} and \textit{irreducible} weighted matrix of
the graph in synchronous\footnote{In a \textit{synchronous} protocol, all nodes activate at the same time and perform communication updates. On the other hand, in an \textit{asynchronous} protocol, each node has its concept of time defined by a local timer, which randomly triggers either by the local timer or by a message from neighboring nodes. The algorithms guaranteed to work with no \textit{a priori} bound on the time for updates are called \textit{totally asynchronous}, and those that need the knowledge of \textit{a priori} bound, known as B-connectivity assumption, are called \textit{partially asynchronous} (see \cite{17} and \cite[Ch. 6-7]{qlearnbertsekas}).}	protocol. We show that state-dependent weighted random operator of the graph is \textit{quasi-nonexpansive}\footnote{It has been shown in \cite{alavianiTAC} that \textit{state-independent} weighted random operator of the graph has \textit{nonexpansivity} property.}; therefore, imposing a priori distribution
of random communication topologies in not required. Thus, it contains
random arbitrary networks with/without asynchronous protocols for more general class of switched networks. As an extension of the the distributed optimization problem, we provide a more general mathematical optimization problem than that defined in \cite{alavianiTAC}, namely minimization of a convex function over the fixed-value point set of a quasi-nonexpansive random operator. Consequently, the reduced optimization problem to distributed optimization includes both state-independent and state-dependent networks over random arbitrary communication graphs with/without asynchronous protocol (see footnote 3). We prove that the discrete-time algorithm proposed in \cite{alavianiTAC} is utilized for \textit{quasi-nonexpansive} random operators (which include nonexpansive random operators as a special case). The algorithm converges both almost surely and in mean square to the global optimal solution of the optimization problem under suitable assumptions. For the distributed optimization problem, the algorithm reduces to a \textit{totally asynchronous} algorithm (see footnote 2). It should be noted that the distributed algorithm\footnote{We require to clarify that the distributed algorithm in this paper is the \textit{randomized version} of the algorithm presented in \cite{alavianiSignalprocessing}. In \cite{alavianiSignalprocessing}, the convergence under deterministic arbitrary switching (see footnote 1) is provided, while we prove here its stochastic convergence (both almost sure and mean square) under random arbitrary switching. Furthermore, quasi-nonexpansivity property of the state-dependent weighted operator of the graph (defined in \cite{alavianiSignalprocessing}) has not been shown in \cite{alavianiSignalprocessing}, whereas we show it here.} is totally asynchronous but not asynchronous due to synchronized diminishing step size. The algorithm is able to converge even if the weighted matrix of the graph is periodic and irreducible under synchronous protocol. We provide a numerical example where there is \textit{distribution dependency} among random arbitrary switching
graphs and apply the distributed algorithm to validate the results, while no existing references can conclude results (see Example 1). \textbf{This version provides proofs, mean square convergence of the proposed algorithm, a numerical example, and larger range of a parameter (i.e., $\beta$) in the algorithm, that have not been presented in the preliminary version (i.e., \cite{alavianiCDC2021})}.

This paper is organized as follows. In Section 2, preliminaries on convex analysis and stochastic convergence are given. In Section 3, formulations of the distributed optimization problem and the mathematical optimization are provided. Algorithm and its convergence analysis are presented in Section 4. Finally, a numerical example is given in order to show advantages of the results in Section 5, followed by conclusions and future work in Section 6.

\textit{Notations:} $\Re$ denotes the set of all real  numbers. For any vector $z \in \Re^{n},\Vert z \Vert_{2}=\sqrt{z^{T}z},$ and for any matrix $Z \in \Re^{n \times n},\Vert Z \Vert_{2}=\sqrt{\lambda_{\max}(Z^{T}Z)}=\sigma_{\max}(Z)$ where $Z^{T}$ represents the transpose of matrix $Z$, $\lambda_{\max}$ represents maximum eigenvalue, and $\sigma_{\max}$ represents largest singular value. Sorted in an increasing order with respect to real parts, $\lambda_{2}(Z)$ represents the second eigenvalue of a matrix $Z$. $Re(r)$ represents the real part of the complex number $r$. For any matrix $Z \in \Re^{n \times n}$ with $Z=[z_{ij}]$,  $\Vert Z \Vert_{1}= \max_{1 \leq j \leq n} \{\sum_{i=1}^{n} \vert z_{ij} \vert \}$ and $\Vert Z \Vert_{\infty}= \max_{1 \leq i \leq n} \{\sum_{j=1}^{n} \vert z_{ij} \vert \}$. $I_{n}$ represents Identity matrix of size $n \times n$ for some $n \in \mathbb{N}$ where $\mathbb{N}$ denotes the set of all natural numbers. $\nabla f(x)$ denotes the gradient of the function $f(x)$. $\otimes$  denotes the Kronecker product. $\times$ represents Cartesian product. $E[x]$ denotes Expectation of the random variable $x$.

\section{Preliminaries}

A vector $v \in \Re^{n}$ is said to be a \textit{stochastic vector} when its components $v_{i}, i=1,2,...,n$, are non-negative and their sum is equal to 1; a square $n \times n$ matrix $V$ is said to be a \textit{stochastic matrix} when each row of $V$ is a stochastic vector. A square $n \times n$ matrix $V$ is said to be \textit{doubly stochastic matrix} when both $V$ and $V^{T}$ are stochastic matrices.

Let $\mathcal{H}$ be a real Hilbert space with norm $\Vert . \Vert$ and inner product $\langle . , . \rangle$. An operator $A: \mathcal{H} \longrightarrow \mathcal{H}$ is said to be \textit{monotone} if $\langle x-y,Ax-Ay \rangle \geq 0$ for all $x,y \in \mathcal{H}$. $A:\mathcal{H} \longrightarrow \mathcal{H}$ is \textit{$\rho$-strongly monotone} if $\langle x-y,Ax-Ay \rangle \geq \rho \Vert x-y \Vert^{2}$ for all $x,y \in \mathcal{H}$. A differentiable function $f:\mathcal{H} \longrightarrow \Re$ is \textit{$\rho$-strongly convex} if $\langle x-y,\nabla f(x)-\nabla f(y) \rangle \geq \rho \Vert x-y \Vert^{2}$ for all $x,y \in \mathcal{H}$. Therefore, a function is $\rho$-strongly convex if its gradient is $\rho$-strongly monotone. A convex differentiable function $f:\mathcal{H} \longrightarrow \Re$ is $\mathcal{L}$\textit{-strongly smooth} if 
$$\langle x-y,\nabla f(x)-\nabla f(y) \rangle \leq \mathcal{L} \Vert x-y \Vert^{2}, \forall x,y \in \mathcal{H}.$$

A mapping $B:\mathcal{H} \longrightarrow \mathcal{H}$ is said to be \textit{$K$-Lipschitz continuous} if there exists a $K >0$ such that $\Vert Bx-By \Vert \leq K \Vert x-y \Vert$ for all $x,y \in \mathcal{H}$. Let $S$ be a nonempty subset of a Hilbert space $\mathcal{H}$ and $Q:S \longrightarrow \mathcal{H}$. The point $x$ is called a \textit{fixed point} of $Q$ if $x=Q(x)$. And, $\text{Fix}(Q)$ denotes the set of all fixed points of $Q$. 

Let $\omega^{*}$ and $\omega$ denote elements in the sets $\Omega^{*}$ and $\Omega$, respectively, where $\Omega=\Omega^{*} \times \Omega^{*} \hdots$. Let $(\Omega^{*},\sigma)$ be a measurable space ($\sigma$-sigma algebra) and $C$ be a nonempty subset of a Hilbert space $\mathcal{H}$. A mapping $x:\Omega^{*} \longrightarrow \mathcal{H}$ is \textit{measurable} if $x^{-1}(U) \in \sigma$ for each open subset $U$ of $\mathcal{H}$. The mapping $T:\Omega^{*} \times C \longrightarrow \mathcal{H}$ is a \textit{random map} if for each fixed $z \in C$, the mapping $T(.,z): \Omega^{*} \longrightarrow \mathcal{H}$ is measurable, and it is \textit{continuous} if for each $\omega^{*} \in \Omega^{*}$ the mapping $T(\omega^{*},.):C \longrightarrow \mathcal{H}$ is continuous.

\begin{definition}
	A measurable mapping $x:\Omega^{*} \longrightarrow C, C \subseteq \mathcal{H},$ is a \textit{random fixed point} of the random map $T:\Omega^{*} \times C \longrightarrow \mathcal{H}$ if $T(\omega^{*},x(\omega^{*}))=x(\omega^{*})$ for each $\omega^{*} \in \Omega^{*}$.
\end{definition}

\begin{definition}
	\cite{alavianiTAC} If there exists a point $\hat{x} \in \mathcal{H}$ such that $\hat{x}=T(\omega^{*},\hat{x})$ for all $\omega^{*} \in \Omega^{*}$, it is called \textit{fixed-value point}, and $FVP(T)$ represents the set of all fixed-value points of $T$.
\end{definition}

\begin{definition}
	Let $C$ be a nonempty subset of a Hilbert space $\mathcal{H}$ and $T:\Omega^{*} \times C \longrightarrow C$ be a random map. The map $T$ is said to be
	
	\textit{1) nonexpansive random operator} if for each $\omega^{*} \in \Omega^{*}$ and for arbitrary $x,y \in C$ we have 
	\begin{equation}\label{bbbbb}
		\Vert T(\omega^{*},x)- T(\omega^{*},y) \Vert \leq \Vert x-y \Vert,
	\end{equation}
	
	\textit{2) quasi-nonexpansive random operator} if for any $x \in C$ we have 
	$$\Vert T(\omega^{*},x)-\xi(\omega^{*}) \Vert \leq \Vert x-\xi(\omega^{*}) \Vert$$
	where $\xi:\Omega^{*} \longrightarrow C$ is a random fixed point of $T$ (see Definition 1).
\end{definition}

Note that if $\Vert T(\omega^{*},x)- T(\omega^{*},x) \Vert \leq \gamma \Vert x-y \Vert, 0 \leq \gamma <1,$ holds in (\ref{bbbbb}), the operator is called \textit{(Banach) contraction}.

\begin{remark}
	If a nonexpansive random operator has a random fixed point, then it is a quasi-nonexpansive random operator. From Definitions 2 and 3, if a quasi-nonexpansive random operator has a fixed-value point, say $x^{*}$, then we have for any $x \in C$ that 
	\begin{equation}\label{bbbbb2}
		\Vert T(\omega^{*},x)-x^{*} \Vert \leq \Vert x-x^{*} \Vert.
	\end{equation} 
\end{remark}

\begin{proposition}
	\cite[Th. 1]{quasi11} If $C$ is a closed convex subset of a Hilbert space $\mathcal{H}$ and $T:C \longrightarrow C$ is quasi-nonexpansive, then $Fix(T)$ is a nonempty closed convex set.
\end{proposition}

\begin{definition}
	A sequence of random variables $x_{t}$ is said to converge 	
	
	\textit{1) pointwise (surely)} to $x$ if for every $\omega \in \Omega$, 
	$$\lim_{t \longrightarrow \infty} \Vert x_{t}(\omega)-x(\omega) \Vert=0,$$
	
	\textit{2) almost surely} to $x$ if there exists a subset $\mathcal{A} \subseteq \Omega$ such that $Pr(\mathcal{A})=0$, and for every $\omega \notin \mathcal{A}$, 
	$$\lim_{t \longrightarrow \infty} \Vert x_{t}(\omega)-x(\omega) \Vert=0,$$
	
	\textit{3) in mean square} to $x$ if 
	$$E[\Vert x_{t}-x \Vert^{2}]\longrightarrow 0 \quad{} as \quad{} t \longrightarrow \infty.$$
\end{definition}

\noindent \textbf{Lemma 1.} \textit{\cite[Ch. 5]{47} Let $W \in \Re^{m \times m}$. Then $\Vert W \Vert_{2} \leq \sqrt{\Vert W \Vert_{1} \Vert W \Vert_{\infty}}.$}

\noindent \textbf{Lemma 2.} \textit{\cite{48} Let $\{ a_{t} \}_{t=0}^{\infty}$ be a sequence of nonnegative real numbers satisfying $a_{t+1} \leq (1-b_{t})a_{t}+b_{t}h_{t}, \quad{t \geq 0}$ where $b_{t} \in [0,1], \sum_{t=0}^{\infty} b_{t}=\infty$, and $\displaystyle \limsup_{t \longrightarrow \infty} h_{t} \leq 0$. Then $\displaystyle \lim_{t \longrightarrow \infty} a_{t}=0$.}

\noindent \textbf{Lemma 3.} \textit{Let the sequence $\{x_{t}\}_{t=0}^{\infty}$ in a real Hilbert space $\mathcal{H}$ be bounded for each realization $\omega \in \Omega$ and converge almost surely to $x^{*}$. Then the sequence converges in mean square to $x^{*}$.}

\noindent \textit{Proof:} See the proof of Theorem 2 in \cite{alavianiTAC}.

The Cucker-Smale weight \cite{38}, which depends on distance between two agents $i$ and $j$, is of the form 
\begin{equation}\label{cuckersmaleweight}
	\mathcal{W}_{ij}(x_{i},x_{j})=\frac{Q}{(\sigma^{2}+\Vert x_{i}-x_{j} \Vert^{2}_{2})^{\beta}}
\end{equation}
where $Q,\sigma>,$ and $\beta \geq 0$.

\section{Problem Formulation}

In social networks, an agent weighs the opinions of others based on how close its opinion (or state) and theirs are, that motivates consideration of \textit{state-dependent networks}. Vehicular platoon can be modeled as both position-dependent (or state-dependent by considering the position as state) and random arbitrary networks, that is a practical example for motivation of this work. Therefore, there are two combined networks: 1) a network induced by states' weights, and 2) the underlying random arbitrary network  (see Section III in the preliminary version, i.e., \cite{alavianiCDC2021}, for details). The combined \textit{state-dependent} \& \textit{random} arbitrary network is formulated as follows.

A network of $m \in \mathbb{N}$ nodes labeled by the set $\mathcal{V}=\lbrace 1,2,...,m \rbrace $ is considered. The topology of the interconnections among nodes is not fixed but defined by  a set of graphs $\mathcal{G}(\omega^{*})=(\mathcal{V},\mathcal{E}(\omega^{*}))$ where $\mathcal{E}(\omega^{*})$ is the ordered edge set $\mathcal{E}(\omega^{*}) \subseteq \mathcal{V} \times \mathcal{V}$ and $\omega^{*} \in \Omega^{*}$ where $\Omega^{*}$ is the set of all possible communication graphs, i.e., $\Omega^{*}=\lbrace \mathcal{G}_{1}, \mathcal{G}_{2}, ..., \mathcal{G}_{\bar{N}}  \rbrace$. We assume that $(\Omega^{*},\sigma)$ is a measurable space where $\sigma$ is the $\sigma$-algebra on $\Omega^{*}$. We write $\mathcal{N}_{i}^{in} (\omega^{*})/\mathcal{N}_{i}^{out} (\omega^{*})$ for the labels of agent $i$'s in/out neighbors at graph $\mathcal{G}(\omega^{*})$ so that there is an arc in $\mathcal{G}(\omega^{*})$ from vertex $j/i$ to vertex $i/j$ only if agent $i$ receives/sends information from/to agent $j$. We write $\mathcal{N}_{i}(\omega^{*})$ when $\mathcal{N}_{i}^{in}(\omega^{*})=\mathcal{N}_{i}^{out}(\omega^{*})$. It is assumed that there are no communication delay or communication noise in the network.

It should be noted that in our formulation, the \textit{in} and \textit{out} neighbors of each agent $i \in \mathcal{V}$ at each graph $\mathcal{G}_{i} \in \Omega^{*},i=1, \hdots, \bar{N},$ are fixed, and we will consider that the weights of links are possibly state-dependent. For instance, an agent pays attention arbitrarily at each time to its friends while it weighs the difference between its opinion and others for decision (see \cite[Sec. III]{alavianiSignalprocessing} for more details).


We associate for each node $i \in \mathcal{V}$ a convex cost function $f_{i}:\Re^{n} \longrightarrow \Re$ which is only observed by node $i$. The objective of each agent is to find a solution of the following optimization problem:
$$\underset{s} \min \sum_{i=1}^{m} f_{i}(s)$$
where $s \in \Re^{n}$. Since each node $i$ knows only its own $f_{i}$, the nodes cannot individually calculate the optimal solution and, therefore, must collaborate to do so.

The above problem can be formulated based on local variables of the agents as

\begin{equation}\label{pprr1}
	\begin{aligned}
		& \underset{x}{\text{min}}
		& & f(x):=\sum_{i=1}^{m} f_{i}(x_{i}) \\
		& \text{subject to}
		& & x_{1}= \hdots =x_{m}
	\end{aligned}
\end{equation}
where $x=[x_{1}^{T},x_{2}^{T}, \hdots, x_{m}^{T}]^{T}, x_{i} \in \Re^{n}, i \in \mathcal{V}$, and the constraint set is reached through \textit{state-dependent} interactions and \textit{random} (arbitrary) communication graphs. The set 
\begin{equation}\label{consensussub}
	\mathcal{C}:=\{ x \in \Re^{mn} | x_{i}=x_{j}, 1 \leq i,j \leq m, x_{i} \in \Re^{n} \}
\end{equation} 
is known as \textit{consensus subspace} which is a convex set. Note that the Hilbert space considered in this paper for the distributed optimization problem is $\mathcal{H}=(\Re^{mn}, \Vert . \Vert_{2}).$

We show $W(\omega^{*},x):=\mathcal{W}(\omega^{*},x) \otimes I_{n}$ and $\mathcal{W}(\omega^{*},x)=[\mathcal{W}_{ij}(\omega^{*},x_{i},x_{j})]$ for the \textit{state-dependent} weighted matrix of the fixed graph $\omega^{*} \in \Omega^{*}$ in a \textit{switching} network having all possible communication topologies in the set $\Omega^{*}$. For instance, if nodes are not activated at some time $\tilde{t}$ for communication updates in asynchronous protocol, and/or there are no edges in occuring graph at the time $\tilde{t}$,  then $\mathcal{W}(\omega^{*}_{\tilde{t}},x_{\tilde{t}})=I_{m}$. 

Now we impose Assumptions 1 and 2 below on $\mathcal{W}(\omega^{*},x)$.

\noindent \textbf{Assumption 1.} \textit{For each fixed $\omega^{*} \in \Omega^{*}$, the weights $\mathcal{W}_{ij}(\omega^{*},x_{i},x_{j}):\Omega^{*} \times \Re^{n} \times \Re^{n} \longrightarrow [0,1]$ are continuous, and the state-dependent weighted matrix of the graph is doubly stochastic for all $\omega^{*} \in \Omega^{*}$, i.e.,}

\textit{\textit{i)} $\sum_{j \in \mathcal{N}_{i}^{in}(\omega^{*}) \cup \{ i \}} \mathcal{W}_{ij}(\omega^{*},x_{i},x_{j})=1, i=1,2,...,m,$}

\textit{\textit{ii)} $\sum_{j \in \mathcal{N}_{i}^{out}(\omega^{*}) \cup \{ i \}} \mathcal{W}_{ij}(\omega^{*},x_{i},x_{j})=1, i=1,2,...,m.$}

\textit{Assumption 1 allows us to remove the \textit{couple} of information
	exchange with the analysis of our proposed algorithm and to
	consider random graphs together.} The state-dependent weight $\mathcal{W}_{ij}(\omega^{*},x_{i},x_{j})$ between any two agents $i$ and $j$ in Assumption 1 is \textit{general} and may be a function of distance or other forms of interactions. Note that any network with undirected links and continuous weights $\mathcal{W}_{ij}(\omega^{*},x_{i},x_{j})$ satisfies Assumption 1 since the weighted matrix of the graph is symmetric (and thus doubly stochastic).

\noindent \textbf{Assumption 2.} \textit{The union of the graphs in $\Omega^{*}$ is strongly connected for all $x \in \Re^{mn}$, i.e., 
\begin{equation}\label{asss2}
	Re[\lambda_{2}(\sum_{\omega^{*} \in \Omega^{*}} (I_{m}-\mathcal{W}(\omega^{*},x)))]>0, \quad{} \forall x \in \Re^{mn}.
\end{equation}}

Assumption 2 guarantees that the information sent from each node is eventually received by every other node. The set $\mathcal{C}$ defined in (\ref{consensussub}) (which is the constraint set of (\ref{pprr1})) can be obtained from the set
\begin{equation}\label{settttttttt11}
	\{x  | W(\omega^{*},x)x=x, \forall \omega^{*} \in \Omega^{*}, \textit{with Assumptions 1 and 2} \}
\end{equation}
(see \cite[Appendices A and B]{alavianiSignalprocessing} by setting $G=\Omega^{*}$ for the proof). This allows us to reformulate (\ref{pprr1}) as
\begin{equation}\label{1}
	\begin{aligned}
		& \underset{x}{\text{min}}
		& & f(x):=\sum_{i=1}^{m} f_{i}(x_{i}) \\
		& \text{subject to}
		& & W(\omega^{*},x)x=x , \forall \omega^{*} \in \Omega^{*}.
	\end{aligned}
\end{equation}
Thus, a solution of (\ref{pprr1}) can be attained by solving (\ref{1}) with Assumptions 1 and 2. 

The random operator $T(\omega^{*},x):=W(\omega^{*},x)x$ is called \textit{state-dependent weighted random operator of the graph} (see \cite[Def. 8]{alavianiTAC}, \cite[Def. 4]{alavianiSignalprocessing}). From Definition 2 and (\ref{settttttttt11}), we have $FVP(T)=\mathcal{C}$ with Assumptions 1 and 2.

Now we show that the random operator $T(\omega^{*},x):=W(\omega^{*},x)x$ with Assumption 1 is \textit{quasi-nonexpansive} in the Hilbert space $\mathcal{H}=(\Re^{mn},\Vert . \Vert_{2})$. Let $z \in FVP(T)=\mathcal{C}$. Since $z \in \mathcal{C}$ and $W(\omega^{*},x)$ is a stochastic matrix (see Assumption 1) for all $\omega^{*} \in \Omega^{*}, x \in \mathcal{H}$, we have $W(\omega^{*},x)z=z$. Therefore, we obtain  
\begin{align*}
	\Vert T(\omega^{*},x)-z \Vert_{2} &=\Vert W(\omega^{*},x)x-W(\omega^{*},x)z \Vert_{2} \\
	&\leq \Vert W(\omega^{*},x) \Vert_{2} \Vert x-z \Vert_{2}.
\end{align*}
Since $W(\omega^{*},x)$ is doubly stochastic by Assumption 1, we have from Lemma 1 (where $\Vert W(\omega^{*},x) \Vert_{1}=\Vert W(\omega^{*},x) \Vert_{\infty}=1$) that $\Vert W(\omega^{*},x) \Vert_{2} \leq 1, \forall \omega^{*} \in \Omega^{*}.$ Hence,
\begin{equation}\label{quasinonexp}
	\Vert T(\omega^{*},x)-z \Vert_{2} \leq \Vert W(\omega^{*},x) \Vert_{2} \Vert x-z \Vert_{2} \leq \Vert x-z \Vert_{2}
\end{equation}
which implies that the random operator $T(\omega^{*},x)$ is \textit{quasi-nonexpansive} (see Remark 1). 

Problem (\ref{1}) is a special case of the \textit{general} class of problem presented in Problem 1 below where $T(\omega^{*},x):=W(\omega^{*},x)x$. It is to be noted that Problem 3 in \cite{alavianiTAC} is defined for \textit{nonexpansive} random operator, while we define Problem 1 below for \textit{quasi-nonexpansive} random operator which contains nonexpansive random operator as a special case (see Remark 1).

\noindent \textbf{Problem 1:} \textit{Let $\mathcal{H}$ be a real Hilbert space. Assume that the problem is feasible, namely $FVP(T) \neq \emptyset$. Given a convex function $f:\mathcal{H} \longrightarrow \Re$ and a quasi-nonexpansive random mapping $T:\Omega^{*} \times \mathcal{H} \longrightarrow \mathcal{H}$, the problem is to find $x^{*} \in \underset{x}{\operatorname{argmin}} f(x)$  such that $x^{*} $ is a fixed-value point of $T(\omega^{*},x)$, i.e., we have the following minimization problem}

\begin{equation}\label{prob1}
	\begin{aligned}
		& \underset{x}{\text{min}} 
		& & f(x)  \\
		& \text{subject to}
		& & x \in FVP(T)
	\end{aligned}
\end{equation}
\textit{where $FVP(T)$ is the set of fixed-value points of the random operator $T(\omega^{*},x)$ (see Definition 2).}

\begin{remark}
A fixed-value point of a quasi-nonexpansive random mapping is a common fixed point of a family of quasi-nonexpansive non-random mappings $T(\omega^{*},.)$ for each $\omega^{*}$. From Preposition 1, the fixed point set of a quasi-nonexpansive non-random mapping $T(\omega^{*},.)$ for each $\omega^{*}$ is a convex set. It is well-known that the intersection of convex sets (finite, countable, or uncountable) is convex. Thus, $FVP(T)$ is a convex set, and Problem 1 is a \textit{convex optimization} problem.
\end{remark}

\section{Algorithm and Its Convergence}

Here, we present that the proposed algorithm in \cite{alavianiTAC} (which works for \textit{nonexpansive} random operators) is applicable for solving Problem 1 with \textit{quasi-nonexpansive} random operators. Thus, we propose the following algorithm for solving Problem 1:
\begin{equation}\label{7}
	x_{t+1}=\alpha_{t} (x_{t}- \beta  \nabla f(x_{t}))+(1-\alpha_{t}) \hat{T}(\omega_{t}^{*},x_{t}),
\end{equation}
where $\hat{T}(\omega_{t}^{*},x_{t}):=(1-\eta) x_{t}+\eta T(\omega_{t}^{*},x_{t}), \eta \in (0,1), \alpha_{t} \in [0,1],$ and $\omega^{*}_{t}$ is a realization of the set $\Omega^{*}$ at time $t$. \textit{The \textit{challenge} of extending the result in \cite{alavianiTAC} is to use weaker property (\ref{bbbbb2}) which is valid for all $x \in \mathcal{H}$ and $x^{*} \in FVP(T)$, instead of stronger property (\ref{bbbbb}) which is valid for all $x,y \in \mathcal{H}$. }

Let $(\Omega^{*}, \sigma)$ be a measurable space where $\Omega^{*}$ and $\sigma$ are defined in Section 3. Consider a probability measure $\mu$ defined on the space $(\Omega,\mathcal{F})$ where 
$$\Omega =\Omega^{*} \times \Omega^{*} \times \Omega^{*} \times \hdots$$
and $\mathcal{F}$ is a sigma algebra on $\Omega$ such that $(\Omega,\mathcal{F},\mu)$ forms a probability space. We denote a realization in this probability space by $\omega \in \Omega$. We have the following assumptions.

\noindent \textbf{Assumption 3.} \textit{$f(x)$ is continuously differentiable,  $\rho$-strongly convex, and $\nabla f(x)$ is $K$-Lipschitz continuous.}

\noindent \textbf{Assumption 4.} \textit{There exists a nonempty subset $\tilde{K} \subseteq \Omega^{*}$ such that $FVP(T)=\{ \tilde{z} | \tilde{z} \in \mathcal{H}, \tilde{z}=T(\bar{\omega},\tilde{z}), \forall \bar{\omega} \in \tilde{K} \}$, and each element of $\tilde{K}$ occurs infinitely often almost surely.}

Assumption 4 is weaker than existing assumptions for random networks as explained in details in Remark 3 below.

\begin{remark}
	\cite{alavianiTAC} If the sequence $\{ \omega^{*}(t) \}_{n=0}^{\infty}$ is mutually independent with $\sum_{t=0}^{\infty} Pr_{t}(\bar{\omega})=\infty$ where $Pr_{t}(\bar{\omega})$ is the probability of (a particular element) $\bar{\omega}$  occurring at time $t$, then Assumption 4 is satisfied. Moreover, any ergodic stationary sequences $\{ \omega^{*}(t) \}_{t=0}^{\infty}, Pr(\bar{\omega})>0,$ satisfy Assumption 4. Consequently, any time-invariant Markov chain with its unique stationary distribution as the initial distribution satisfies Assumption 4. 
\end{remark}

\subsection{Almost Sure Convergence}

Before we give our theorems, we need to extend Lemma 5 in \cite{alavianiTAC} (which is for \textit{nonexpansive} random operators) to \textit{quasi-nonexpansive} random operators. Hence, we have the following lemma.

\noindent \textbf{Lemma 4.} \textit{Let $\mathcal{H}$ be a real Hilbert space, $\hat{T}(\omega^{*},x):=(1-\eta)x+\eta T(\omega^{*},x), \omega^{*} \in \Omega^{*}, x \in \mathcal{H},$ with a quasi-nonexpansive random operator $T$, $FVP(T) \neq \emptyset$, and $\eta \in (0,1]$. Then}

\textit{\textit{(i)} $FVP(T)=FVP(\hat{T}).$}

\textit{\textit{(ii)} $\langle x-\hat{T}(\omega^{*},x),x-z \rangle \geq \frac{\eta}{2} \Vert x-T(\omega^{*},x) \Vert^{2}, \forall z \in \quad{} \quad{} \quad{} \quad{} FVP(T), \forall \omega^{*} \in \Omega^{*}.$}

\textit{\textit{(iii)} $\hat{T}(\omega^{*},x)$ is quasi-nonexpasnive.}

\noindent \textit{Proof.} See Appendix A.

We present the main theorem in this paper as follows.

\begin{theorem}
Consider Problem 1 with Assumptions 3 and 4. Let $\beta \in (0,\frac{2}{K})$ and $\alpha_{t} \in [0,1], t \in \mathbb{N} \cup \lbrace 0 \rbrace$ such that 

\textit{(a)} $\displaystyle \lim_{t \longrightarrow \infty} \alpha_{t}=0,$ 

\textit{(b)} $\sum_{t=0}^{\infty} \alpha_{t}=\infty.$ 

Then starting from any initial point, the sequence generated by (\ref{7}) globally converges almost surely to the unique solution of the problem.
\end{theorem}

Note that the range of $\beta$ in Theorem 1 in \cite{alavianiCDC2021} (i.e., the preliminary version of this paper) is $\beta \in (0,\frac{2 \rho}{K^{2}})$ which is enlarged to $\beta \in (0,\frac{2}{K})$ in Theorem 1 above. This is due to the fact that according to definitions of strong convexity and strong smoothness of a differentiable convex function $f$ (see also parts (5)-(6) in \cite[p. 38]{63nnnnn}), we always have $\rho \leq K$. Hence, $\frac{2 \rho}{K^{2}} \leq \frac{2}{K}$. An advantageous of this enlargement is to have more choice to select the parameter $\beta.$ An example of $\alpha_{t}$ satisfying \textit{(a)} and \textit{(b)} in Theorem 1 is $\alpha_{t} :=\frac{1}{(1+t)^{\zeta}}$ where $\zeta \in (0,1]$.

\begin{remark}	
	As seen from the proof of Theorem 1, an advantage of the proposed technique in \cite{alavianiTAC} (and thus here) is that we are able to analyze stochastic processes in a \textit{fully deterministic} way (see Remark 12 in \cite{alavianiTAC} for details). 
\end{remark}

\noindent \textit{Proof of Theorem 1.} We prove Theorem 1 in three steps, i.e.

\textit{Step 1:} $\{ x_{t} \}_{t=0}^{\infty}, \forall \omega \in \Omega,$ is bounded (see Lemma 5 in Appendix B).

\textit{Step 2:} $\{ x_{t} \}_{t=0}^{\infty}$ converges almost surely to a random variable supported by the feasible set (see Lemma 6 in Appendix C).

\textit{Step 3:} $\{ x_{t} \}_{t=0}^{\infty}$ converges almost surely to the optimal solution (see Lemma 7 in Appendix D).

\subsection{Mean Square Convergence}

Due to the fact that almost sure convergence in general does not imply mean square convergence and vice versa, we show the mean square convergence of the random sequence generated by Algorithm (\ref{7}) in the following theorem.

\begin{theorem}	
	Consider Problem 1 with Assumptions 3 and 4. Suppose that $\beta \in (0,\frac{2 }{K})$ and $\alpha_{t} \in [0,1], t \in \mathbb{N} \cup \lbrace 0 \rbrace$, satisfies \textit{(a)} and \textit{(b)} in Theorem 1. Then starting from any initial point, the sequence generated by (\ref{7}) globally converges in mean square to the unique solution of the problem. 
\end{theorem}

\noindent \textit{Proof.} From \textit{Step 1}, Theorem 1, and Lemma 3, one can prove Theorem 2.

\subsection{Distributed Optimization}

Distributed optimization problem with \textit{state-dependent} interactions over \textit{random} arbitrary networks is a special case of Problem 1 (see Section 3). Hence, Algorithm (\ref{7}) is directly applied to solve (\ref{1}) in a distributed manner under the consideration that each $f_{i}(x_{i})$ is $\rho$-strongly convex and $\nabla f_{i}(x_{i})$ is $K$-Lipschitz. Thus, we give the following corollary of Theorems 1 and 2.

\noindent \textbf{Corollary 1.} \textit{Consider the optimization (\ref{1}) with Assumptions 1, 2, and 4. Assume that each $f_{i}(x_{i})$ is $\rho$-strongly convex and $\nabla f_{i}(x_{i})$ is $K$-Lipschitz for $i=1, \hdots,m$. Suppose that $\beta \in (0,\frac{2}{K}), \eta \in (0,1),$ and $\alpha_{t} \in [0,1], t \in \mathbb{N} \cup \lbrace 0 \rbrace$ satisfies \textit{(a)} and \textit{(b)} in Theorem 1. Then starting from any initial point, the sequence generated by the following \textit{distributed} algorithm based on local information for each agent $i$}
\begin{align}\label{77777}
	x_{i,t+1} &=\alpha_{t} (x_{i,t}- \beta  \nabla f_{i}(x_{i,t}))+(1-\alpha_{t}) ((1-\eta) x_{i,t} \nonumber \\
	&\quad{}+\eta \sum_{j \in \mathcal{N}_{i}^{in}(\omega^{*}_{t}) \cup \{ i \}} \mathcal{W}_{ij}(\omega^{*}_{t},x_{i},x_{j})x_{j,t}),
\end{align}
\textit{globally converges both almost surely and in mean square to the unique solution of the problem.}

Algorithm (\ref{77777}) is \textit{totally asynchronous} algorithm (see footnote 2) without requiring a priori distribution or B-connectivity (see footnote 2) of switched graphs. B-connectivity assumption satisfies Assumption 4. The algorithm is \textit{not} asynchronous due to synchronized diminishing step size $\alpha_{t}$. The algorithm still works in the case where state-dependent/state-independent weighted matrix of the graph is \textit{periodic} and \textit{irreducible} in synchronous protocol. Detailed properties of Algorithm (\ref{77777}) for \textit{time-varying} (see footnote 1) networks has been studied in \cite{alavianiSignalprocessing} and can be induced for random networks (see also footnote 4).

\begin{remark}
The convergence rate of
a totally asynchronous algorithm in general cannot be established. Determining rate of convergence of (\ref{77777}) based on suitable assumptions is left for future work. An asynchronous and totally asynchronous algorithm for distributed optimization over random networks with \textit{state-independent} interactions has recently been proposed in \cite{alavianiAcc2022first}. As a special case of distributed optimization over state-independent networks, asynchronous and total asynchronous algorithms have been given for average consensus and solving linear algebraic equations in \cite{alavianiACC2019} and \cite{alavianiLAE}, respectively (see \cite[Sec. I]{alavianiAcc2022first} for details). 
\end{remark}

\section{Numerical Example}

We give a practical example of distributed optimization with state-dependent interactions of Cucker-Smale form \cite{38} with random (arbitrary) communication links where there are \textit{distribution dependencies} among random arbitrary switched graphs. We mention that the following example has been solved over \textit{time-varying} (see footnote 1) networks in \cite{alavianiSignalprocessing}, while we solve it here over \textit{random} arbitrary networks, where there are \textit{distribution dependency} among switched communication graphs, to show the capability of Algorithm (\ref{77777}).

\noindent \textbf{Example 1.} \textit{(Distributed Optimization over Random Arbitrary Networks for an Automated Warehouse):} Consider $m$ robots on the shop floor in a warehouse. Assigning tasks to robotic agents is modeled as optimization problems in an automated warehouse \cite{68aaa}, that are solved by a centralized processor and are neither scalable nor can handle autonomous entities \cite{68aaa}. Moreover, due to large number of robots, the robots must handle tasks in collaborative manner \cite{68aaa} due to computational restriction of a centralized processor. If we assume that the communications among robots are carried out via a wireless network, then the signal power at a receiver is inversely proportional to some power of the distance between transmitter and receiver \cite{69aaa}. Therefore, if we consider the position as the state for each robot, then the weights of the links between robots are \textit{state-dependent}.

\begin{figure}[t!]
	\centering
	\includegraphics[width=0.9\textwidth]{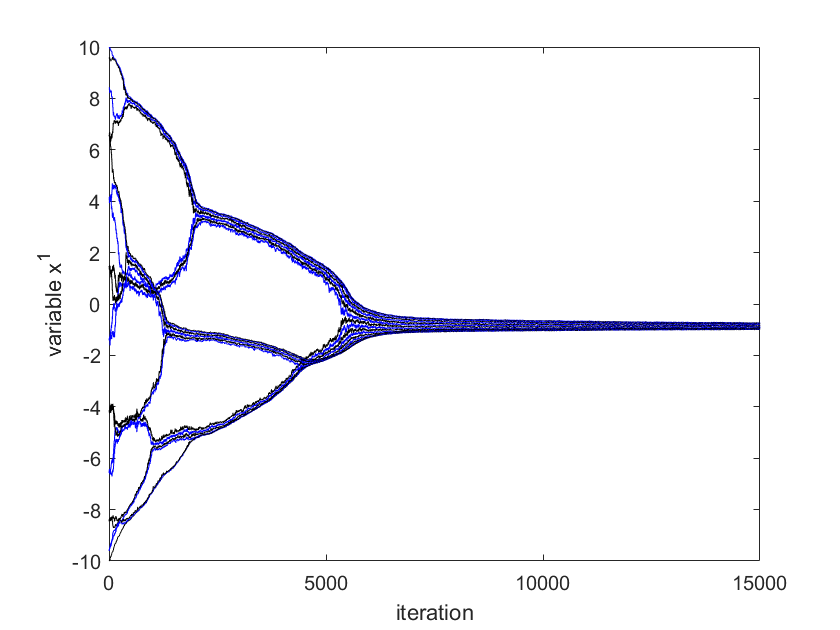}
	\caption{Variables $x_{i}^{1}, i=1, \hdots,20,$ of the robotic agents with weights of the form (\ref{cuckerexample1}). This figure shows that the variables are getting consensus when the robots communicate for one realization of random network with distribution dependency.}\label{fig1}
\end{figure}

Assume that $m=20$ robots bring some loads from different initial places to a place for delivery. The desired place to put the loads is determined to minimize the pre-defined cost as sum of squared distances to the initial places of the robots as	 
\begin{equation}\label{optimExampl1}
	\underset{s} \min \sum_{i=1}^{20} \Vert s-d_{i} \Vert^{2}_{2}
\end{equation}
where $s \in \Re^{2}$ is the decision variable, and $d_{i}$ is the position of the initial place of the load $i$ on the two-dimensional shop floor. The above problem is reformulated as the following problem based on the local variables of the agents:

\begin{equation*}
	\begin{aligned}
		& \underset{x}{\text{min}}
		& & f(x):=\sum_{i=1}^{20} 0.5 \Vert x_{i}-d_{i} \Vert^{2}_{2}  \\
		& \text{subject to}
		& & x_{1}=x_{2}=\hdots=x_{20}
	\end{aligned}
\end{equation*}
where $x_{i}=[x_{i}^{1},x_{i}^{2}]^{T}$, and the constraint set is reached via \textit{distance-dependent} network with \textit{random} communication graphs. 

\begin{figure}[t!]
	\centering
	\includegraphics[width=0.9\textwidth]{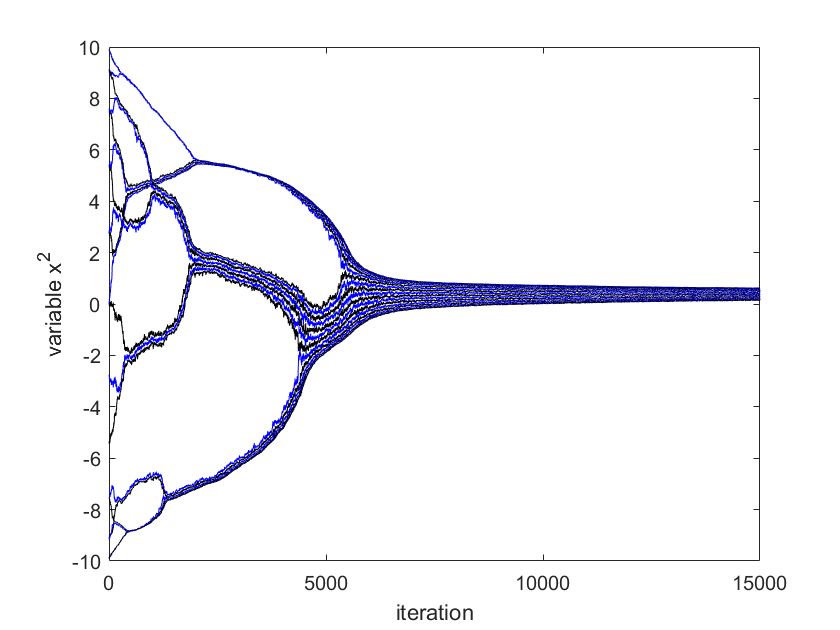}
	\caption{Variables $x_{i}^{2}, i=1, \hdots,20,$ of the robotic agents with weights of the form (\ref{cuckerexample1}). This figure shows that the variables are getting consensus when the robots communicate for one realization of random network with distribution dependency.}\label{fig2}
\end{figure}

The topology of the underlying undirected graph is assumed to be a line graph, i.e., $1 \longleftrightarrow 2 \hdots \longleftrightarrow 20$, for minimal connectivity among robots. Based on weighing property of wireless communication network mentioned eralier, the weight of the link between robots $i$ and $j$ is modeled to be of Cucker-Smale form (see Section 2)
\begin{equation}\label{cuckerexample1}
	\mathcal{W}_{ij}(x_{i},x_{j})= \frac{0.25}{1+\Vert x_{i}-x_{j} \Vert^{2}_{2}}.
\end{equation} 
One can see that the weight of each link at each time $t$ is only determined by the states of the agents, and hence no local property is assumed or determined a priori for all $t$ in Algorithm (\ref{77777}) (see \cite{alavianiSignalprocessing} for details). It is easy to check that $f_{i}(x_{i}):=0.5 \Vert x_{i}-d_{i} \Vert^{2}_{2}, i=1,2,...,20,$ are 1-strongly convex, and $\nabla f_{i}(x_{i})$ are $1$-Lipschitz continuous.

\begin{figure}[t!]
	\centering
	\includegraphics[width=0.9\textwidth]{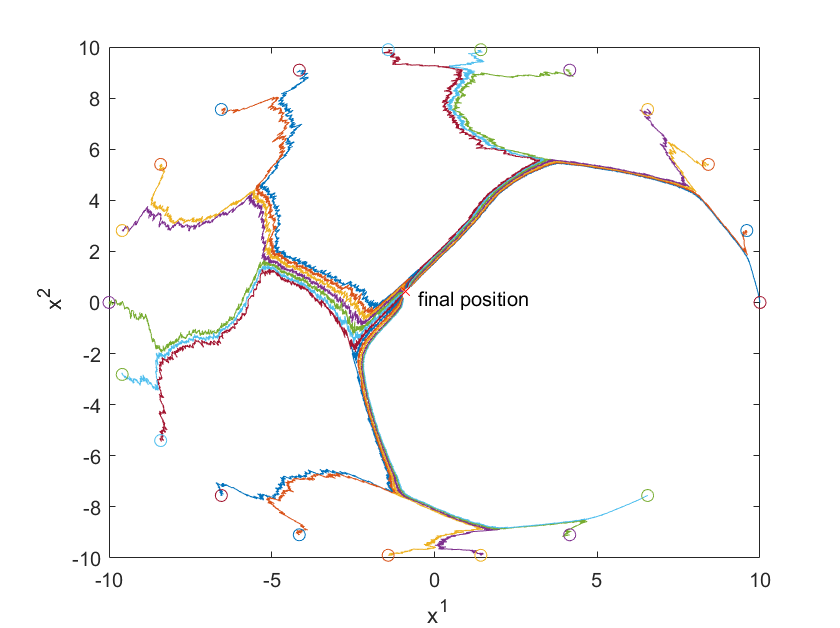}
	\caption{Two-dimensional (2D) plot of variables $x^{1}$ and $x^{2}$, in Figures \ref{fig1} and \ref{fig2}, where the initial positions of agents are shown with 'o', and the final position is shown with 'x'.}\label{fig3}
\end{figure}

\begin{figure}[t!]
	\centering
	\includegraphics[width=0.9\textwidth]{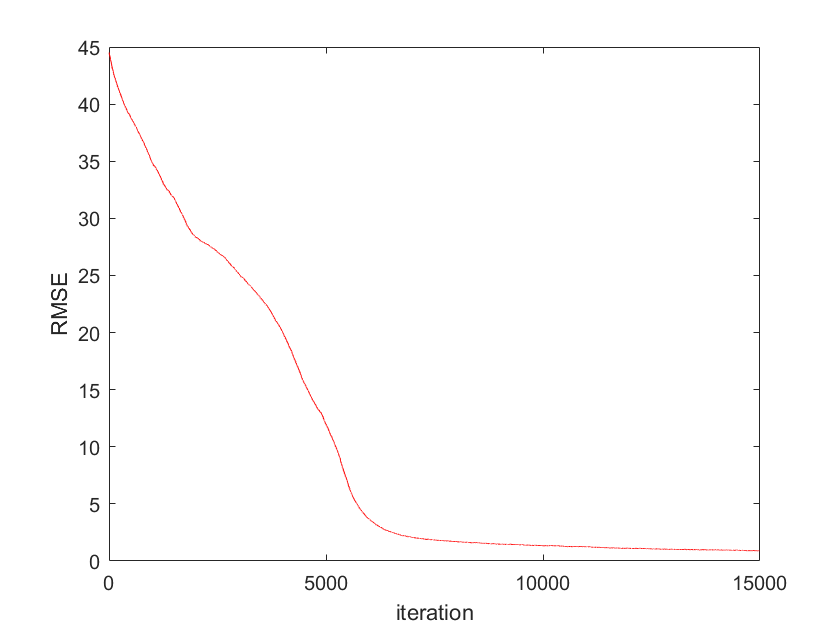}
	\caption{The error in Example 1 with weights of the form (\ref{cuckerexample1}) for one realization of the random network with distribution dependency.}\label{figerror}
\end{figure}

We consider that each link has independent and identically distributed (i.i.d.) Bernoulli distribution with $Pr(failure)=0.5$ in every $\tilde{N}$-interval, and at the iteration $k \tilde{N}, k=1, \hdots,$ the link that has worked the minimum number of the times in the previous $\tilde{N}$-interval occurs. If some links have the same number of occurrences in the previous $\tilde{N}$-interval, then one is chosen randomly. Here, we have the graphs $\mathcal{\omega}^{*} \in \Omega^{*}=\{\mathcal{G}_{1}, \hdots, \mathcal{G}_{19}\}$. Thus the sequence $\{\omega^{*}_{t}\}_{t=0}^{\infty}$ is not independent. It has been shown in \cite{alavianiTAC} that the each graph $\mathcal{G}_{i}, i=1, \hdots, 19,$ occurs infinitely often almost surely. Moreover, the union of the graphs is strongly connected for all $x \in \Re^{40}$. Therefore, Assumption 4 is fulfilled. Thus the conditions of Theorems 1 and 2 are satisfied.

We use $\eta=0.8, \alpha_{t}=\frac{1}{1+t}, t \geq 0, \beta=\frac{1}{K}=\dfrac{1}{1}$ for simulation. The initial position of agent $i$ is chosen to be $x_{i,0}=[10 cos(\frac{(i-1)2 \pi}{22}),10 sin(\frac{(i-1)2 \pi}{22})]^{T}$. The optimal solution of (\ref{optimExampl1}) in centralized way can be computed as mean of $d_{i}, i=1,\hdots,20,$ and is $s^{*}=[-0.9002,0.4111]^{T}$. The results given by Algorithm (\ref{77777}) are shown in Figure \ref{fig1}-\ref{figerror}. The error $e_{t}:=\Vert x_{t}-s^{*} \otimes \textbf{1}_{20} \Vert_{2}$, where $x_{t}=[x_{1,t}^{T},\hdots,x_{20,t}^{T}]^{T},$    is given in Fig. \ref{figerror}. The two-dimensional (2D) plot is shown in Figure \ref{fig3}. Figures \ref{fig1}-\ref{figerror} show that the positions of robotic agents are approaching the solution of the optimization (\ref{optimExampl1}) for one realization of random network with \textit{distribution dependency}. Note that no existing result can solve this problem since the weights of links are both \textit{position-dependent} and \textit{randomly} arbitrarily activated.

\begin{figure}[h]
	\centering
	\includegraphics[width=0.9\textwidth]{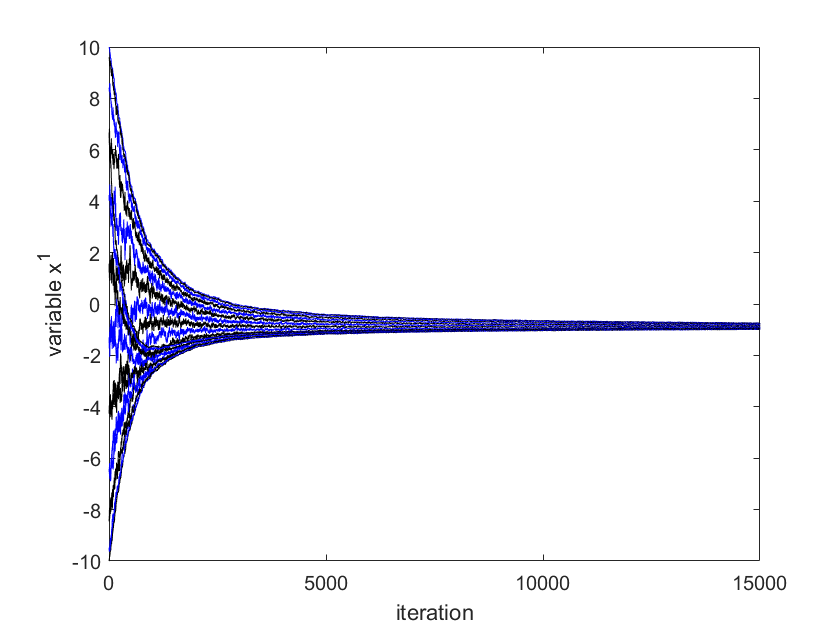}
	\caption{Variables $x_{i}^{1}, i=1, \hdots,20,$ of the robotic agents with weights of the form (\ref{weightlog}). This figure shows that the variables are getting consensus when the robots communicate for one realization of random network with distribution dependency.}\label{fig1log}
\end{figure}

\begin{figure}[h]
	\centering
	\includegraphics[width=0.9\textwidth]{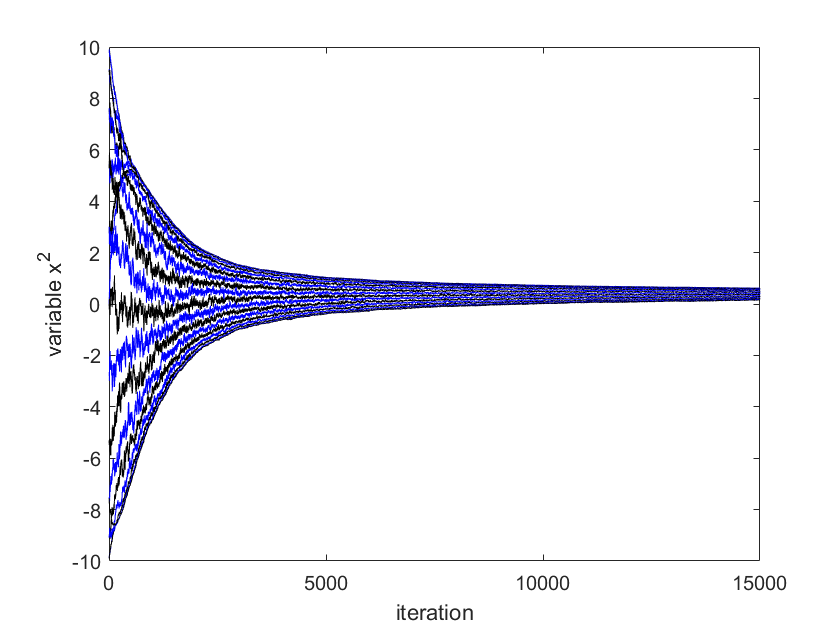}
	\caption{Variables $x_{i}^{2}, i=1, \hdots,20,$ of the robotic agents with weights of the form (\ref{weightlog}). This figure shows that the variables are getting consensus when the robots communicate for one realization of random network with distribution dependency.}\label{fig2log}
\end{figure}

We also simulate the above example with different weights than Cucker-Smale form, i.e.,

\begin{equation}\label{weightlog}
	\mathcal{W}_{ij}(x_{i},x_{j})=\frac{0.25}{1+log^{2}(1+\Vert x_{i}-x_{j} \Vert_{2})},
\end{equation}
and the results are shown in Figures \ref{fig1log}-\ref{fig2log}. The figures show that the variables of agents are getting consensus on the optimal solution of the problem.

\section{Conclusions and Future Work}
Distributed optimization with both state-dependent interactions and random (arbitrary)
networks is considered. It is shown that the state-dependent weighted random operator of the graph is quasi-nonexpansive; thus, it is not required to impose a priori distribution of random communication topologies on switching graphs. A more general optimization problem than that addressed in the literature is provided. A gradient-based discrete-time algorithm using diminishing step size is provided that is able to converge both almost surely and in mean square to the global solution of the optimization problem under suitable assumptions. Moreover, it reduces to a totally asynchronous algorithm for the distributed optimization problem. Relaxing strong convexity assumption of cost functions and/or doubly stochasticity assumption of communication graphs opens problems for future research.

\begin{appendices}

\section{}

\noindent \textit{Proof of Lemma 4.} 	

\textit{(i)} The proof is the same as the proof of part \textit{(i)} of Lemma 5 in \cite{alavianiTAC}.

\textit{(ii)} We have from quasi-nonexpansivity of $T(\omega^{*},x)$ for arbitrary $x \in \mathcal{H}$ that
\begin{equation}\label{ii1}
	\Vert T(\omega^{*},x)-z \Vert^{2} \leq \Vert x-z \Vert^{2}, \forall z \in FVP(T),\forall \omega^{*} \in \Omega^{*}.
\end{equation}
In a Hilbert space $\mathcal{H}$, we have
\begin{equation}\label{hilbertinequ}
	\Vert u+v \Vert^{2}=\Vert u \Vert^{2}+\Vert v \Vert^{2}+2 \langle u,v \rangle, \forall u,v \in \mathcal{H}.
\end{equation}
From (\ref{hilbertinequ}), we obtain for all $z \in FVP(T)$ and for all $\omega^{*} \in \Omega^{*}$ that
\begin{align}
	\Vert T(\omega^{*},x)-z \Vert^{2} &=\Vert T(\omega^{*},x)-x+x-z \Vert^{2} \nonumber\\
	&=\Vert T(\omega^{*},x)-x \Vert^{2}+\Vert x-z \Vert^{2} \nonumber\\
	&\quad +2 \langle T(\omega^{*},x)-x,x-z \rangle.\label{ii2}                          
\end{align}
Substituting (\ref{ii2}) for (\ref{ii1}) yields
\begin{equation}\label{ii3}
	2 \langle x-T(\omega^{*},x),x-z \rangle \geq \Vert T(\omega^{*},x)-x \Vert^{2}. 
\end{equation}
From the definition of $\hat{T}(\omega_{t}^{*},x_{t})$ (see (\ref{7})), substituting $x-T(\omega^{*},x)=\frac{x-\hat{T}(\omega^{*},x)}{\eta}$ for the left hand side of the inequality (\ref{ii3}) implies \textit{(ii)}. Thus the proof of part \textit{(ii)} of Lemma 4 is complete.

\textit{(iii)} We have from quasi-nonexpansivity of $T(\omega^{*},x)$ for $z \in FVP(T)$ and arbitrary $x \in \mathcal{H}$ that
\begin{align*}
	\Vert \hat{T}(\omega^{*},x)-z \Vert  &\leq (1-\eta) \Vert x-z \Vert+\eta \Vert T(\omega^{*},x)-z \Vert \\
	&\leq (1-\eta) \Vert x-z \Vert+\eta \Vert x-z \Vert \\
	&= \Vert x-z \Vert, \forall \omega^{*} \in \Omega^{*}.
\end{align*}
Therefore, $\hat{T}(\omega^{*},x)$ is a quasi-nonexpansive random operator, and the proof of part \textit{(iii)} of Lemma 4 is complete.

\section{}

\noindent \textbf{Lemma 5.} \textit{The sequence $\{ x_{t} \}_{t=0}^{\infty}, \forall \omega \in \Omega,$ generated by (\ref{7}) is bounded with Assumption 3. }

\noindent \textit{Proof.} Since the cost function is smooth and strongly convex and the constraint set is nonempty and closed, the problem has a unique solution. Let $x^{*}$ be the unique solution of the problem. We can write $x^{*}=\alpha_{t} x^{*} +(1-\alpha_{t})x^{*}, \forall t \in \mathbb{N} \cup \{ 0 \}$. Therefore, we have
\begin{align*}
	\Vert x_{t+1}-x^{*} \Vert &=\Vert \alpha_{t} (x_{t}- \beta \nabla f(x_{t}))   +(1-\alpha_{t}) \hat{T}(\omega_{t}^{*},x_{t})-x^{*} \Vert \\
	&= \Vert \alpha_{t}(x_{t}-\beta \nabla f(x_{t})-x^{*})  +(1-\alpha_{t})(\hat{T}(\omega_{t}^{*},x_{t})-x^{*}) \Vert \\
	&\leq \alpha_{t}  \Vert x_{t}-\beta \nabla f(x_{t})-x^{*} \Vert  +(1-\alpha_{t})  \Vert \hat{T}(\omega_{t}^{*},x_{t})-x^{*} \Vert.
\end{align*}
Since $x^{*}$ is the solution, we have that $x^{*} \in FVP(T)=FVP(\hat{T})$ (see part \textit{(i)} of Lemma 4). Due to the fact that $\hat{T}(\omega^{*},x)$ is a quasi-nonexpansive random operator (see part \textit{(iii)} of Lemma 4), the above can be written as
\begin{align}
	\Vert x_{t+1}-x^{*} \Vert &\leq \alpha_{t}  \Vert x_{t}-\beta \nabla f(x_{t})-x^{*} \Vert  +(1-\alpha_{t})  \Vert \hat{T}(\omega_{t}^{*},x_{t})-x^{*} \Vert \nonumber \\
	& \leq \alpha_{t}  \Vert x_{t}-\beta \nabla f(x_{t})-x^{*} \Vert  +(1-\alpha_{t})  \Vert x_{t}-x^{*} \Vert. \label{9}
\end{align}

Since $\nabla f_{i}(x_{i})$ is $K$-Lipschitz, $f_{i}(x_{i})$ is $K$-strongly smooth  (see \cite[Lem. 3.4]{62nnnnn}). When $f_{i}(x_{i})$ is $\rho$-strongly convex and $K$-strongly smooth, the operator $H(x):=x-\beta \nabla f(x)$ where $\beta \in (0,\frac{2}{K})$ is a contraction (see \cite[p. 15]{63nnnnn} for details). Indeed, there exists a $0< \gamma \leq 1$ such that 
\begin{equation}\label{10}
	\Vert x-y-\beta(\nabla f(x)-\nabla f(y)) \Vert \leq (1-\gamma) \Vert x-y \Vert, \forall x,y \in \mathcal{H}. 
\end{equation}
We have that
\begin{align}
	 \Vert x_{t}-\beta \nabla f(x_{t})-x^{*} \Vert   
	&= \Vert x_{t}-x^{*}-\beta(\nabla f(x_{t})-\nabla f(x^{*}))-\beta \nabla f(x^{*}) \Vert  \nonumber \\
	&\leq \Vert x_{t}-x^{*}-\beta(\nabla f(x_{t})-\nabla f(x^{*})) \Vert +\beta  \Vert \nabla f(x^{*}) \Vert. \label{beta11} 
\end{align}
Therefore, (\ref{10}) and (\ref{beta11}) imply
\begin{align}
	 \Vert x_{t}-\beta \nabla f(x_{t})-x^{*} \Vert 
	&\leq \Vert x_{t}-x^{*}-\beta(\nabla f(x_{t})-\nabla f(x^{*})) \Vert  +\beta  \Vert \nabla f(x^{*}) \Vert \nonumber \\
	&\leq (1-\gamma)  \Vert x_{t}-x^{*} \Vert  + \beta  \Vert \nabla f(x^{*}) \Vert. \label{11}
\end{align}
Substituting (\ref{11}) for (\ref{9}) yields
\begin{align*}
	\Vert x_{t+1}-x^{*} \Vert &\leq (1-\gamma \alpha_{t})  \Vert x_{t}-x^{*} \Vert +\alpha_{t} \beta \Vert \nabla f(x^{*}) \Vert \\
	&= (1-\gamma \alpha_{t})  \Vert x_{t}-x^{*} \Vert + \gamma \alpha_{t}(\frac{\beta \Vert \nabla f(x^{*}) \Vert}{\gamma})
\end{align*}
which by induction implies that
$$ \Vert x_{t+1}-x^{*} \Vert \leq max \lbrace  \Vert x_{0}-x^{*} \Vert, \frac{\beta \Vert \nabla f(x^{*}) \Vert}{\gamma}  \rbrace$$
that implies $ \Vert x_{t}-x^{*} \Vert, t \in \mathbb{N} \cup \lbrace 0 \rbrace, \forall \omega \in \Omega$, is bounded. Therefore, $\{ x_{t} \}_{t=0}^{\infty}$ is bounded for all $\omega \in \Omega$.

\section{}

\noindent \textit{Lemma 6.} \textit{The sequence $\{ x_{t} \}_{t=0}^{\infty}$ generated by (\ref{7}) converges almost surely to a random variable supported by the feasible set.}

\noindent \textit{Proof.} From (\ref{7}) and $x_{t}=\alpha_{t} x_{t}+(1-\alpha_{t}) x_{t}$, we have
\begin{equation}\label{march0}
	x_{t+1}-x_{t}+\alpha_{t} \beta  \nabla f(x_{t})=(1-\alpha_{t}) (\hat{T}(\omega_{t}^{*},x_{t})-x_{t}),
\end{equation}
and thus
\begin{align}
	 \langle x_{t+1}-x_{t}+\alpha_{t} \beta  \nabla f(x_{t}),x_{t}-x^{*} \rangle 
	=-(1-\alpha_{t})  \langle x_{t}-\hat{T}(\omega_{t}^{*},x_{t}),x_{t}-x^{*} \rangle. \label{march1}
\end{align}
Due to $x^{*} \in FVP(T)$, we have from part \textit{(ii)} of Lemma 4 that
\begin{equation}\label{march2}
	\langle x_{t}-\hat{T}(\omega_{t}^{*},x_{t}),x_{t}-x^{*} \rangle \geq \frac{\eta}{2} \Vert x_{t}-T(\omega_{t}^{*},x_{t}) \Vert^{2}.
\end{equation}
We get from (\ref{march1}) and (\ref{march2}) that
\begin{align}
	 \langle x_{t+1}-x_{t}+\alpha_{t} \beta  \nabla f(x_{t}),x_{t}-x^{*} \rangle  \leq - \frac{\eta}{2} (1-\alpha_{t}) \Vert x_{t}-T(\omega_{t}^{*},x_{t}) \Vert^{2} \label{march3}
\end{align}
or 
\begin{align}
	- \langle x_{t}-x_{t+1},x_{t}-x^{*} \rangle \leq -\alpha_{t} \langle \beta \nabla f(x_{t}),x_{t}-x^{*} \rangle-  \frac{\eta}{2} (1-\alpha_{t}) \Vert x_{t}-T(\omega_{t}^{*},x_{t}) \Vert^{2}. \label{march4}
\end{align}
In a Hilbert space $\mathcal{H},$ we have for any $u,v \in \mathcal{H}$ that
\begin{equation}\label{march5}
	\langle u,v \rangle =-\frac{1}{2} \Vert u-v \Vert^{2}+\frac{1}{2} \Vert u \Vert^{2}+\frac{1}{2} \Vert v \Vert^{2}.
\end{equation}
We obtain from (\ref{march5}) that
\begin{equation}\label{march6}
	\langle x_{t}-x_{t+1},x_{t}-x^{*}\rangle=-C_{t+1}+C_{t}+\frac{1}{2} \Vert x_{t}-x_{t+1} \Vert^{2}
\end{equation}
where $C_{t}:=\frac{1}{2} \Vert x_{t}-x^{*} \Vert^{2}$. We get from (\ref{march4}) and (\ref{march6}) that
\begin{align}
	 C_{t+1}-C_{t}-\frac{1}{2} \Vert x_{t}-x_{t+1} \Vert^{2} 
	&\leq - \alpha_{t} \langle \beta \nabla f(x_{t}),x_{t}-x^{*} \rangle  -\frac{\eta}{2} (1-\alpha_{t}) \Vert x_{t}-T(\omega_{t}^{*},x_{t}) \Vert^{2}. \label{march7} 
\end{align}
From (\ref{march0}) and (\ref{hilbertinequ}) we obtain
\begin{align}
	 \Vert x_{t+1}-x_{t} \Vert^{2}   
	&= \Vert - \alpha_{t} \beta \nabla f(x_{t})+(1-\alpha_{t}) (\hat{T}(\omega^{*}_{t},x_{t})-x_{t}) \Vert^{2}  \nonumber \\
	&= \alpha_{t}^{2}  \Vert \beta \nabla f(x_{t}) \Vert^{2}+ (1-\alpha_{t})^{2} \Vert \hat{T}(\omega^{*}_{t},x_{t})-x_{t} \Vert^{2} \nonumber\\
	&\quad - 2 \alpha_{t} (1-\alpha_{t}) \langle \beta \nabla f(x_{t}),\hat{T}(\omega_{t}^{*},x_{t})-x_{t} \rangle.  \label{march9} 
\end{align}
We know that $\Vert \hat{T}(\omega^{*}_{t},x_{t})-x_{t} \Vert=\eta \Vert x_{t}-T(\omega^{*}_{t},x_{t}) \Vert$. Since $\alpha_{t} \in [0,1]$, we have also that $(1-\alpha_{t})^{2} \leq (1-\alpha_{t})$. Using these facts and multiplying both sides of (\ref{march9}) by $\frac{1}{2}$ yield
\begin{align}
	 \frac{1}{2} \Vert x_{t+1}-x_{t} \Vert^{2}  
	&= \frac{1}{2} \alpha_{t}^{2}  \Vert \beta \nabla f(x_{t}) \Vert^{2}    +\frac{1}{2} (1-\alpha_{t})^{2} \eta^{2} \Vert T(\omega^{*}_{t},x_{t})-x_{t} \Vert^{2} \nonumber \\
	&\quad - \alpha_{t} (1-\alpha_{t}) \langle \beta \nabla f(x_{t}),\hat{T}(\omega_{t}^{*},x_{t})-x_{t} \rangle  \nonumber \\
	&\leq \frac{1}{2} \alpha_{t}^{2}  \Vert \beta \nabla f(x_{t}) \Vert^{2}    +\frac{1}{2} (1-\alpha_{t}) \eta^{2} \Vert T(\omega^{*}_{t},x_{t})-x_{t} \Vert^{2} \nonumber \\
	&\quad - \alpha_{t} (1-\alpha_{t}) \langle \beta \nabla f(x_{t}),\hat{T}(\omega_{t}^{*},x_{t})-x_{t} \rangle.  \label{march10}
\end{align}
We obtain from (\ref{march7}) and (\ref{march10}) that
\begin{align}
	C_{t+1}-C_{t} &\leq \frac{1}{2} \Vert x_{t+1}-x_{t} \Vert^{2}  - \alpha_{t} \langle \beta \nabla f(x_{t}),x_{t}-x^{*} \rangle \nonumber\\
	&\quad -\frac{\eta}{2} (1-\alpha_{t}) \Vert x_{t}-T(\omega_{t}^{*},x_{t}) \Vert^{2} \nonumber\\
	& \leq -(\frac{1}{2}-\frac{\eta}{2}) \eta  (1-\alpha_{t}) \Vert x_{t}-T(\omega_{t}^{*},x_{t}) \Vert^{2} \nonumber\\
	&\quad + \alpha_{t}(\frac{1}{2}\alpha_{t} \Vert \beta \nabla f(x_{t}) \Vert^{2} \nonumber\\
	&\quad -\langle \beta \nabla f(x_{t}),x_{t}-x^{*} \rangle \nonumber\\
	&\quad -(1-\alpha_{t}) \langle \beta \nabla f(x_{t}),\hat{T}(\omega_{t}^{*},x_{t})-x_{t} \rangle). \label{march11}  
\end{align}
We claim that there exists $t_{0} \in \mathbb{N}$ such that the sequence $\{ C_{t} \}$ is non-increasing for $t \geq t_{0}$. We use proof by contradiction and assume that this is not true. Then there exists a subsequence $\{ C_{t_{j}} \}$ such that $C_{t_{j}+1}-C_{t_{j}}>0$ which with (\ref{march11}) implies
\begin{align}
	0 &<C_{t_{j}+1}-C_{t_{j}} \nonumber\\
	&\leq -(\frac{1}{2}-\frac{\eta}{2}) \eta  (1-\alpha_{t_{j}}) \Vert x_{t_{j}}-T(\omega_{t_{j}}^{*},x_{t_{j}}) \Vert^{2} \nonumber\\
	& \quad{} + \alpha_{t_{j}}(\frac{1}{2} \alpha_{t_{j}} \beta^{2}\Vert \nabla f(x_{t_{j}}) \Vert^{2}  -\langle \beta \nabla f(x_{t_{j}}),x_{t_{j}}-x^{*} \rangle \nonumber\\
	& \quad{}-(1-\alpha_{t_{j}}) \langle \beta \nabla f(x_{t_{j}}),\hat{T}(\omega_{t_{j}}^{*},x_{t_{j}})-x_{t_{j}} \rangle). \label{march12}    
\end{align}
Since $ \{ x_{t} \}$ is bounded, $\nabla f(x)$ is continuous, and $\eta \in (0,1)$,  we get from (\ref{march12}) and Theorem 1 \textit{(a)} that
\begin{align}
	0 &< \liminf_{j \longrightarrow \infty} [-(\frac{1}{2}-\frac{\eta}{2}) \eta  (1-\alpha_{t_{j}}) \Vert x_{t_{j}}-T(\omega_{t_{j}}^{*},x_{t_{j}}) \Vert^{2} \nonumber \\
	&\quad +\alpha_{t_{j}}(\frac{1}{2} \alpha_{t_{j}} \Vert \beta \nabla f(x_{t_{j}}) \Vert^{2}  -\langle \beta \nabla f(x_{t_{j}}),x_{t_{j}}-x^{*} \rangle \nonumber \\
	&\quad -(1-\alpha_{t_{j}}) \langle \beta \nabla f(x_{t_{j}}),\hat{T}(\omega_{t_{j}}^{*},x_{t_{j}})-x_{t_{j}} \rangle)]  \leq 0
\end{align}
that is a contradiction. Hence, there exists $t_{0} \in \mathbb{N}$ such that the sequence $\{ C_{t} \}$ is non-increasing for $n \geq t_{0}$. Since $\{ C_{t} \}$ is bounded below, it converges for all $\omega \in \Omega$.

Now we take the limit of both sides of (\ref{march11}) and utilize the convergence of $\{ C_{t} \}$, continuity of $\nabla f(x)$, \textit{Step 1}, $\eta \in (0,1)$, and Theorem 1 \textit{(a)} to obtain 
$$\lim_{t \longrightarrow \infty} \Vert x_{t}-T(\omega_{t}^{*},x_{t}) \Vert=0, \quad{} \textit{pointwise (surely)}$$  
which implies that $\{ x_{t} \}_{t=0}^{\infty}$ converges for each $\omega \in \Omega$ since $FVP(T) \neq \emptyset$. Moreover, this together with Assumption 4 implies that $\{ x_{t} \}$ converges almost surely to a random variable supported by $FVP(T)$.

\section{}

\noindent \textbf{Lemma 7.} \textit{The sequence $\{ x_{t} \}_{t=0}^{\infty}$ generated by (\ref{7}) converges almost surely to the optimal solution.}

\noindent \textit{Proof.} Here we prove that $\{ x_{t} \}_{t=0}^{\infty}$ converges almost surely to the optimal solution. Since $x^{*} \in FVP(T)$ is the optimal solution, we have
\begin{equation}\label{optimality}
	\langle \bar{x}-x^{*},\nabla f(x^{*}) \rangle \geq 0, \forall \bar{x} \in FVP(T).
\end{equation}

From (\ref{hilbertinequ}), we have that 
\begin{align}
	 \Vert x_{t+1}-x^{*} \Vert^{2}   
	&= \Vert x_{t+1}-x^{*}+\alpha_{t} \beta \nabla f(x^{*})-\alpha_{t} \beta \nabla f(x^{*}) \Vert^{2} \nonumber \\
	&= \Vert x_{t+1}-x^{*}+\alpha_{t} \beta \nabla f(x^{*}) \Vert^{2}+ \alpha_{t}^{2}  \Vert \beta \nabla f(x^{*}) \Vert^{2} \nonumber\\
	&\quad -2 \alpha_{t}   \langle \beta \nabla f(x^{*}),x_{t+1}-x^{*}+\alpha_{t} \beta \nabla f(x^{*}) \rangle.  \label{19}
\end{align}
We have that $x^{*}=\alpha_{t}x^{*}+(1-\alpha_{t})x^{*}, \forall n \in \mathbb{N} \cup  \{ 0 \}$; We get from this fact and (\ref{7}) that
\begin{align}
	 \Vert x_{t+1}-x^{*}+\alpha_{t} \beta \nabla f(x^{*}) \Vert^{2}   
	&=\Vert \alpha_{t}[x_{t}-x^{*}-\beta( \nabla f(x_{t})- \nabla f(x^{*}))] \nonumber \\
	&\quad +(1-\alpha_{t})[\hat{T}(\omega_{t}^{*},x_{t})-x^{*}] \Vert^{2}.  \label{20}
\end{align}
Furthermore, we have
\begin{align}
	 \langle \beta \nabla f(x^{*}),x_{t+1}-x^{*}+\alpha_{t} \beta \nabla f(x^{*}) \rangle  
	&= \langle \beta \nabla f(x^{*}),x_{t+1}-x^{*} \rangle  +\alpha_{t} \langle \beta \nabla f(x^{*}),\beta \nabla f(x^{*}) \rangle \nonumber \\
	&= \langle \beta \nabla f(x^{*}),x_{t+1}-x^{*} \rangle  +\alpha_{t}  \Vert \beta \nabla f(x^{*}) \Vert^{2}. \label{21}
\end{align}
Substituting (\ref{20}) and (\ref{21}) for (\ref{19}) implies
\begin{align*}
	 \Vert x_{t+1}-x^{*} \Vert^{2} 
	&=\Vert x_{t+1}-x^{*}+\alpha_{t} \beta \nabla f(x^{*}) \Vert^{2}  + \alpha_{t}^{2}  \Vert \beta \nabla f(x^{*}) \Vert^{2} \nonumber\\
	&\quad -2 \alpha_{t}  \langle \beta \nabla f(x^{*}),x_{t+1}-x^{*}+\alpha_{t} \beta \nabla f(x^{*}) \rangle \nonumber\\
	&=\Vert \alpha_{t}[x_{t}-x^{*}-\beta(\nabla f(x_{t})-\nabla f(x^{*}))] \nonumber\\
	&\quad +(1-\alpha_{t})[\hat{T}(\omega_{t}^{*},x_{t})-x^{*}] \Vert^{2} \nonumber\\
	&\quad -2 \alpha_{t} \langle \beta \nabla f(x^{*}),x_{t+1}-x^{*} \rangle - \alpha_{t}^{2} \Vert \beta \nabla f(x^{*}) \Vert^{2} \nonumber\\
	&= \alpha_{t}^{2} \Vert x_{t}-x^{*}-\beta(\nabla f(x_{t})-\nabla f(x^{*})) \Vert^{2} \nonumber\\
	&\quad +(1-\alpha_{t})^{2} \Vert \hat{T}(\omega_{t}^{*},x_{t})-x^{*} \Vert^{2} \nonumber\\
	&\quad +2 \alpha_{t}(1-\alpha_{t}) \langle x_{t}-x^{*} \nonumber\\
	&\quad -\beta (\nabla f(x_{t})- \nabla f(x^{*})),\hat{T}(\omega_{t}^{*},x_{t})-x^{*} \rangle \nonumber\\
	&\quad -2 \alpha_{t}  \langle \beta \nabla f(x^{*}),x_{t+1}-x^{*} \rangle - \alpha_{t}^{2}  \Vert \beta \nabla f(x^{*}) \Vert^{2}. 
\end{align*}
From (\ref{10}), quasi-nonexpansivity property of $\hat{T}(\omega^{*},x)$, and Cauchy--Schwarz inequality, we have
\begin{align}
	\langle x_{t}-x^{*}-\beta (\nabla f(x_{t})- \nabla f(x^{*})),\hat{T}(\omega_{t}^{*},x_{t})-x^{*} \rangle \leq (1-\gamma) \Vert x_{t}-x^{*} \Vert^{2}. \label{caushy}
\end{align}
We get from (\ref{10}) that
\begin{equation}\label{pprrov1}
	\Vert x_{t}-x^{*}-\beta(\nabla f(x_{t})-\nabla f(x^{*})) \Vert^{2} \leq (1-\gamma)^{2} \Vert x_{t}-x^{*} \Vert^{2}.
\end{equation}
We obtain from (\ref{caushy}), (\ref{pprrov1}), and quasi-nonexpansivity property of $\hat{T}(\omega^{*},x)$ that
\begin{align*}
	 \Vert x_{t+1}-x^{*} \Vert^{2} 
	&=\alpha_{t}^{2} \Vert x_{t}-x^{*}-\beta(\nabla f(x_{t})-\nabla f(x^{*})) \Vert^{2} \nonumber\\
	&\quad +(1-\alpha_{t})^{2} \Vert \hat{T}(\omega_{t}^{*},x_{t})-x^{*} \Vert^{2} \nonumber\\
	&\quad +2 \alpha_{t}(1-\alpha_{t}) \langle x_{t}-x^{*}-\beta (\nabla f(x_{t})- \nabla f(x^{*})),  \hat{T}(\omega_{t}^{*},x_{t})-x^{*} \rangle \nonumber\\
	&\quad -2 \alpha_{t}  \langle \beta \nabla f(x^{*}),x_{t+1}-x^{*} \rangle - \alpha_{t}^{2}  \Vert \beta \nabla f(x^{*}) \Vert^{2} \nonumber\\
	&\leq (1-2\gamma \alpha_{t}) \Vert x_{t}-x^{*} \Vert^{2} \nonumber\\
	&\quad +\alpha_{t} (\gamma^{2} \alpha_{t} \Vert x_{t}-x^{*} \Vert^{2}-2  \langle \beta \nabla f(x^{*}),x_{t+1}-x^{*} \rangle) \nonumber\\
	&=(1-\gamma \alpha_{t}) \Vert x_{t}-x^{*} \Vert^{2}-\gamma \alpha_{t} \Vert x_{t}-x^{*} \Vert^{2} \nonumber\\
	&\quad +\alpha_{t}(\gamma^{2} \alpha_{t} \Vert x_{t}-x^{*} \Vert^{2}-2  \langle \beta \nabla f(x^{*}),x_{t+1}-x^{*} \rangle).  
\end{align*}
We obtain from $\gamma \alpha_{t} \Vert x_{t}-x^{*} \Vert^{2} \geq 0$ that
\begin{align*}
	&\quad (1-\gamma \alpha_{t}) \Vert x_{t}-x^{*} \Vert^{2}-\gamma \alpha_{t} \Vert x_{t}-x^{*} \Vert^{2} \nonumber\\
	&\quad +\alpha_{t}(\gamma^{2} \alpha_{t} \Vert x_{t}-x^{*} \Vert^{2}-2  \langle \beta \nabla f(x^{*}),x_{t+1}-x^{*} \rangle) \nonumber\\
	&\leq( 1-\gamma \alpha_{t}) \Vert x_{t}-x^{*} \Vert^{2} \nonumber\\
	&\quad +\alpha_{t}(\gamma^{2} \alpha_{t} \Vert x_{t}-x^{*} \Vert^{2}-2  \langle \beta \nabla f(x^{*}),x_{t+1}-x^{*} \rangle) 
\end{align*}
or finally 
\begin{align}
	\Vert x_{t+1}-x^{*} \Vert^{2} \leq (1-\gamma \alpha_{t}) \Vert x_{t}-x^{*} \Vert^{2}+    \gamma \alpha_{t} (\frac{\gamma^{2} \alpha_{t} \Vert x_{t}-x^{*} \Vert^{2}-2  \langle \beta \nabla f(x^{*}),x_{t+1}-x^{*} \rangle}{\gamma}).  \label{llppp}
\end{align}
From \textit{Step 1}, \textit{Step 2}, (\ref{optimality}), and the condition in Theorem 1 \textit{(a)}, we get 
\begin{align}
	\displaystyle \lim_{t \longrightarrow \infty} (\gamma^{2} \alpha_{t} \Vert x_{t}-x^{*} \Vert^{2}-2 \beta \langle \nabla f(x^{*}),x_{t+1}-x^{*} \rangle)   \leq 0 \quad{\textit{almost surely}}.  \label{lasssstt}
\end{align}
Setting $a_{t}, b_{t}, h_{t}$ in Lemma 2 as 
\begin{align*}
	&a_{t}=\Vert x_{t}-x^{*} \Vert^{2}, \\
	&b_{t}=\gamma \alpha_{t}, \\
	&h_{t}=(\frac{\gamma^{2} \alpha_{t} \Vert x_{t}-x^{*} \Vert^{2}-2 \beta \langle \nabla f(x^{*}),x_{t+1}-x^{*} \rangle}{\gamma}),
\end{align*}
we get from (\ref{llppp}), (\ref{lasssstt}), and the condition in Theorem 1 \textit{(b)} that 
$$\displaystyle \lim_{t \longrightarrow \infty} \Vert x_{t}-x^{*} \Vert^{2}=0 \quad{\textit{almost surely}}.$$ Therefore, $\{ x_{t} \}_{t=0}^{\infty}$ converges almost surely to $x^{*}$.

\end{appendices}











\begin{thebibliography}{xx}

	
	
	
	
	




\bibitem {survey1} Jakoveti\'{c}, D., Bajovi\'{c}, D., Xavier, J., and Moura, J. M. F.: Primal-dual methods for large-scale and distributed convex optimization and data analysis. Proceedings of The IEEE. 108, 1923--1938 (2020)

\bibitem {survey2} Yang, T., Yi, X., Wu, J., Yuan, Y., Wu, D., Meng, Z., Hong, Y., Wang, H., Lin, Z., and Johansson, K. H.: A survey of distributed optimization. Annual Reviews in Control. 47, 278--305 (2019)

\bibitem {survey3} Mazlum, D. K., D\"{o}rfler, F., Sandberg, H., Low, S. H., Chakrabarti, S., Baldick, R., and Lavaei, J.: A survey of distributed optimization and control algorithms for electric power systems. IEEE Trans. on Smart Grid. 8, 2941--2962 (2017)


\bibitem {survey4} Nedi\'{c}, A.: Distributed optimization. Encyclopedia of Systems and Control. 1--12 (2014)


\bibitem{11222} Liberzon, D.: Switching in Systems and Control. Springer, New York (2003)




\bibitem {38} Cucker, F., and Smale, S.: Emergent behavior in flocks. IEEE Trans. Automatic Control. 52, 852--862 (2007)


\bibitem {36} Krause, U.: A discrete nonlinear and non-autonomous model of consensus formation. Communications in Difference Equations. Gordon and Breach, Amsterdam. 227--236 (2000)


\bibitem {37} Totsch, S., and Tadmor, E.: Heterophilious dynamics enhances consensus. SIAM Review. 56, 577-621 (2014)


\bibitem {ozdaglar1} Acemoglu, D., Ozdaglar, A., and Parandehgheibi, A.: Spread of (mis)information in social networks. Games and Economic Behavior. 70, 194--227 (2010)


\bibitem {ozdaglar2} Acemoglu, D., and Ozdaglar, A.: Opinion dynamics and learning in social networks. Dynamic Games and Applications. 1, 3--49 (2011)


\bibitem {ozdaglar3} Acemoglu, D., Como, G., Fagnani, F., and Ozdaglar, A.: Opinion fluctuations and disagreement in social networks. Mathematics of Operations Research. 38, 1--27 (2013)


\bibitem {ozdaglar4} Acemoglu, D., Bimpikis, K., and Ozdaglar, A.: Dynamics of information exchange in endogenous social networks. Theoretical Economics. 9, 41--97 (2014)

\bibitem {opini11111} Heyselmann, R., and Krause, U.: Opinion dynamics and bounded confidence moels, analysis, and simulation. J. Artificial Societies and Social Simulation. 5, 1--33 (2002)


\bibitem {opini22222} Blondel, V. D.,  Hendrickx, J. M., and Tsitsiklis, J. N.: On Krause's multi-agent consensus model with state-dependent connectivity. IEEE Trans. on Autom. Contr. 54, 2586--2597 (2009)


\bibitem {ozdaglar5} Acemoglu, D., Ozdaglar, A., and Yildiz, E.: Diffusion of innovations in social networks. Proc. of 50th IEEE Conf. on Dec. Cont. and Europ. Cont. Conf., Dec. 12-15, Orlando, FL, USA. 2329--2334 (2011)




\bibitem {33} Simonetto, A., Kevicsky, T., and Babu\v{s}ka, R.: Constrained distributed algebraic connectivity maximization in robotic networks. Automatica. 49, 1348--1357 (2013)



\bibitem {34} Kim, Y., and Mesbahi, M.: On maximizing the second smallest eigenvalue of a state-dependent graph laplacian. IEEE Trans. on Autom. Contr. 51, 116--120 (2006)

\bibitem {35} Siljak, D. D.: Dynamic graphs. Nonlin. Analysis: Hybrid Syst. 2, 544--567 (2008)


\bibitem {42} Trianni, V., Simone, D. D., Reina, A., and Baronchelli, A.: Emergence of consensus in a multi-robot network: from abstract models to empirical validation. IEEE Robotics and Automation Letters. 1, 348--353 (2016)


\bibitem {8} Lobel, I., Ozdaglar, A., and Feiger, D.: Distributed multi-agent optimization with state-dependent communication. Math. Program. Ser. B. 129, 255--284 (2011)






\bibitem {statedependentbothways} Jing, G., Zheng, Y., and Wang, L.: Consensus of multiagent systems with distance-dependent communication networks. IEEE Trans. on Neural Networks and Learning Systems. 28, 2712--2726 (2017)


\bibitem {statedependentregional} Jing, G., and Wang, L.: Finite time coordination under state-dependent communication graphs with inherent links. IEEE Trans. on Circuits and Systems-II: Express Briefs. 66, 968--972 (2019)

\bibitem {statedependentnew1111111} Shang, Y.: Constrained consensus in state-dependent directed multiagent networks. IEEE Trans. on Network Science and Engineering. 9, 4416--4425 (2022)



\bibitem {43} Slu\u{c}iak, O., and Rupp, M.: Consensus algorithm with state-dependent weights. IEEE Trans. Signal Processing. 64, 1972--1985 (2016)

\bibitem {44} Bogojeska, A., Mirchev, M., Mishkovski, I., and Kocarev, L.: Synchronization and consensus in state-dependent networks. IEEE Trans. on Circuits and Systems-I: Regular Papers. 61, 522--529 (2014)

\bibitem {45} Awad, A., Chapman, A., Schoof, E., Narang-Siddarth, A., and Mesbahi, M.: Time-scale separation on networks: consensus, tracking, and state-dependent interactions. IEEE 54th Annual Conf. on Decision and Control, Dec. 15-18, Osaka, Japan. 6172--6177 (2015)


\bibitem {11} Shi, G., Johansson, K. H., and Hong, Y.: Reaching an optimal consensus: dynamical systems that compute intersections of convex sets. IEEE Trans. Automatic Control. 58, 610--622 (2013)

\bibitem {statedependentnew222222} Verma, A., Vasconcelos, M., Mitra, U., and Touri, B.: Maximal dissent: a state-dependent way to agree in distributed convex optimization. IEEE Trans. on Control of Network Systems. 10, 1783--1795 (2023)


\bibitem {alavianiSignalprocessing} Alaviani, S. Sh., and Elia, N.: Distributed convex optimization with state-dependent (social) interactions and time-varying topologies. IEEE Trans. Signal Processing. 69, 2611-2624 (2021)



\bibitem {alavianiTAC} Alaviani, S. Sh., and Elia, N.: Distributed multiagent convex optimization over random digraphs. IEEE Trans. Automatic Control. 65, 986--998 (2020)

\bibitem {alavianiCDC2021} Alaviani, S. Sh., and Kelkar, A. G.: Distributed convex optimization with state-dependent interactions over random networks. Proc. of IEEE Conf. on Decision and Control, Dec. 13-17, Austin, Texas, USA. 3149--3153 (2021)

\bibitem{17}
Tsitsiklis, J. N.: Problems in decentralized decision making and computation. \hskip 1em plus
0.5em minus 0.4em\relax Ph.D. dissertation. Dep. Elect. Eng. Comp. Sci., MIT, Cambridge, MA (1984)

\bibitem{qlearnbertsekas}
Bertsekas, D. P., and Tsitsiklis, J. N.: Parallel and Distributed Computation: Numerical Methods. Prentice Hall, Englewood Cliffs (1989)

\bibitem {quasi11}  Dotson, W. G.: Fixed points of quasi-nonexpansive mappings. J. Austral. Math. Soc. 13, 167--170 (1972)

\bibitem {47} Horn, R. A., and Johnson, C. R.: Matrix Analysis.  Cambridge University Press, New York (1985)






\bibitem {48} Xu, H. K.: Iterative algorithms for nonlinear operators. J. London Math. Soc. 66, 240-256 (2002)






\bibitem {alavianiAcc2022first} Alaviani, S. Sh., and Kelkar, A. G.: Asynchronous Algorithms for Distributed Consensus-Based
Optimization and Distributed Resource Allocation over Random
Networks. Proc. of Amer. Cont. Conf., June 8-10, Atlanta, GA, USA. 216--221 (2022)




\bibitem{alavianiACC2019} Alaviani, S. Sh., and Elia, N.: Distributed average consensus over random networks. Proc. of Amer. Cont. Conf., July 10-12, Philadelphia, PA, USA,. 1854--1859 (2019)

\bibitem{alavianiLAE} Alaviani, S. Sh., and Elia, N.: A distributed algorithm for solving linear algebraic equations over random networks. IEEE Trans. on Automatic Control. 66, 2399--2406 (2021)




\bibitem {68aaa} Kattepur, A., Rath, H. K., Mukherjee, A., and Simha, A.: Distributed optimization framework for Industry 4.0 automated warehouses. EAI Endorsed Transactions on Industrial Networks and Intelligent Systems.  5, 1--10 (2018)

\bibitem {69aaa} Pahlavan, K., and Levesque, H.: Wireless Information Networks. Wiley, New York (1995)

\bibitem {62nnnnn} Bubeck, S.: Convex optimization: algorithms and complexity. Foundations and Trends in Machine Learning. 8, 231--357 (2015)

\bibitem {63nnnnn} Ryu, E. K., and Boyd, S.: A primer on monotone operator methods. Appl. Comput. Math. 15, 3--43 (2016)
	
	
	
\end{thebibliography}
\end{document}